\documentclass[usenatbib]{basi}
\usepackage[T1]{fontenc}
\usepackage[british]{babel}
\usepackage[varg]{txfonts}
\usepackage{rotating}
\usepackage{dcolumn}
\begin{document}
\title[Tracing merger history in radio]{Tracing the merger history of MACS clusters using the GMRT\thanks{By
            Surajit Paul, email: \texttt{surajit@physics.unipune.ac.in}, or see
            \texttt{http://www.iucaa.ernet.in/\string~surajit/}}}
\author[S. Paul et~al.]%
       {S. Paul$^{1,2}$\thanks{email: \texttt{surajit@physics.unipune.ac.in}},
       A. Datta$^{3}$ and H. T. Intema $^4$\\
       $^1$IUCAA, Pune, Pune 411007\\
       $^2$Department of Physics, University of Pune, Pune 411007\\
       $^3$CASA, Department of Astrophysical and Planetary Science, University of Colorado, Boulder, CO 80309 \\
       $^4$National Radio Astronomy Observatory, 1003 Lopezville Road, Socorro, NM 87801-0387, USA}

\pubyear{2014}
\volume{00}
\pagerange{\pageref{firstpage}--\pageref{lastpage}}

\date{Received --- ; accepted ---}

\maketitle
\label{firstpage}
\begin{abstract}
Galaxy-cluster merger shocks efficiently accelerate the ambient electrons through diffusive shock acceleration and amplify magnetic field by compressing the Inter Cluster Medium. As a result, such objects produce a significant amount of synchrotron radio emission. Radio halo and Mpc scale peripheral radio relics thus help us to trace back the cluster formation history. To study the dynamical state of the galaxy clusters and their relation to the merging scenario, as a pilot project, we thus observed four suitable candidates from the MACS (Massive Cluster Survey) cluster sample. Observations were carried out simultaneously at 610 and 235 MHz with the GMRT. We observed a rare phenomena in the cluster MACSJ0014.3-3022, which hosts both a peripheral spectacular radio relic and an unusually flat spectrum central giant radio halo of more than 1.5 Mpc dimension. We also report that another cluster MACSJ0152.5-2852 is possibly showing the sign of one of the earliest and young merging system where, we saw a faint 0.5 Mpc radio relic at the cluster periphery. 
\end{abstract}

\vspace{-0.5cm}
\section{Introduction}\label{s:intro}
Before attaining virialization, galaxy clusters passes through the phases of continuous accretion and serial mergers of bigger and bigger galaxy groups. Particularly, mergers are extremely energetic process and the energy released during merger is dissipated in the Inter Cluster Medium (ICM) by thermalizing it through strong collision-less shocks. These strong shocks with an efficient Fermi acceleration of energetic particles could generate strong MHD waves in the upstream and downstream regions of shocks and strongly amplify the upstream magnetic field present in the ICM \citep{bykov2008}. This also transforms kinetic energy to turbulent energy by injecting volume-filling turbulence in the ICM \citep{Subra2006}. Thermal to turbulent energy fraction in such an incident can reach as high as 30$\%$ \citep{Vazza2006,Paul2011}. Turbulent dissipation in such system acts on a significantly longer time-scale than shocks (shock time scale $\sim$ 2 Gyr \citep{Paul2011}), and in principle can stochastically re-accelerate the ambient electrons \citep{Brunetti2007}. As a result, such objects produce a significant amount of synchrotron radio emission. And, thus they are promising candidates for radio observations. Observations of galaxy clusters in radio waves are thus important as they became instrumental in tracing back the formation history of the galaxy-clusters.

Large scale ($l \gtrsim500$ kpc) diffuse radio emission from galaxy clusters are broadly of two types.`Radio halo'  and the `radio relics' of sizes $\sim$ Mpc. Radio halos have smooth morphologies, are extended with sizes $\gtrsim$1 Mpc, unpolarized, and are found at the centres of clusters, co-spatial with the thermal X-ray emitting gas of the ICM. Giant radio relics are observed in the cluster periphery, sometimes showing symmetric or ring-like structures and are highly polarized (p$\sim$ 10-$50 \%$ at 1.5 GHz). They are probably signatures of electrons accelerated at large-scale shocks. The vast majority of clusters with diffuse extended radio sources are massive, X-ray luminous and show signs of undergoing mergers.
\vspace{-0.3cm}
\section{GMRT observations and data analysis}

The principal goal of this work is to obtain the GMRT deep radio maps of diffuse emission from the massive clusters. We want to understand the effect of mergers on production of non-thermal emission and controlling the energy budget of the central and cluster peripheral ICM. It was then an obvious choice to search for the biggest clusters from the MAssive Cluster Survey (MACS). This survey was designed to find the population of strongly evolving clusters, with the most X-ray luminous systems using a specific X-ray selection function described in \citep{Ebeling2001}. From the MACS list (upto 2010; \citet{Ebeling2001,Horesh2010}) we choose the clusters that are showing clear merging activity in X-ray/Temperature map and also from mass distribution calculated from weak lensing of these objects \citep{Zitrin2011}. Significantly, hitherto unexplored faint diffuse radio emission (of radio-halo and/or relic type) is also identified from the 1.4 GHz NVSS survey in their central or peripheral regions. 

In this project, we observed four MACS clusters using GMRT dual band 235 \& 610 MHz with 32 MHz band width. Observation was done during June-August 2011 (Project Code : 20$\_$062,  PI  S. Paul), each with on source time  of 5 hours. For data analysis and imaging, the code 'SPAM' has been used (for details see \citet{Intema2009}). 
\vspace{-0.3cm}
\section{Results}
\begin{figure}[h!]
  \includegraphics[width=4.3cm]{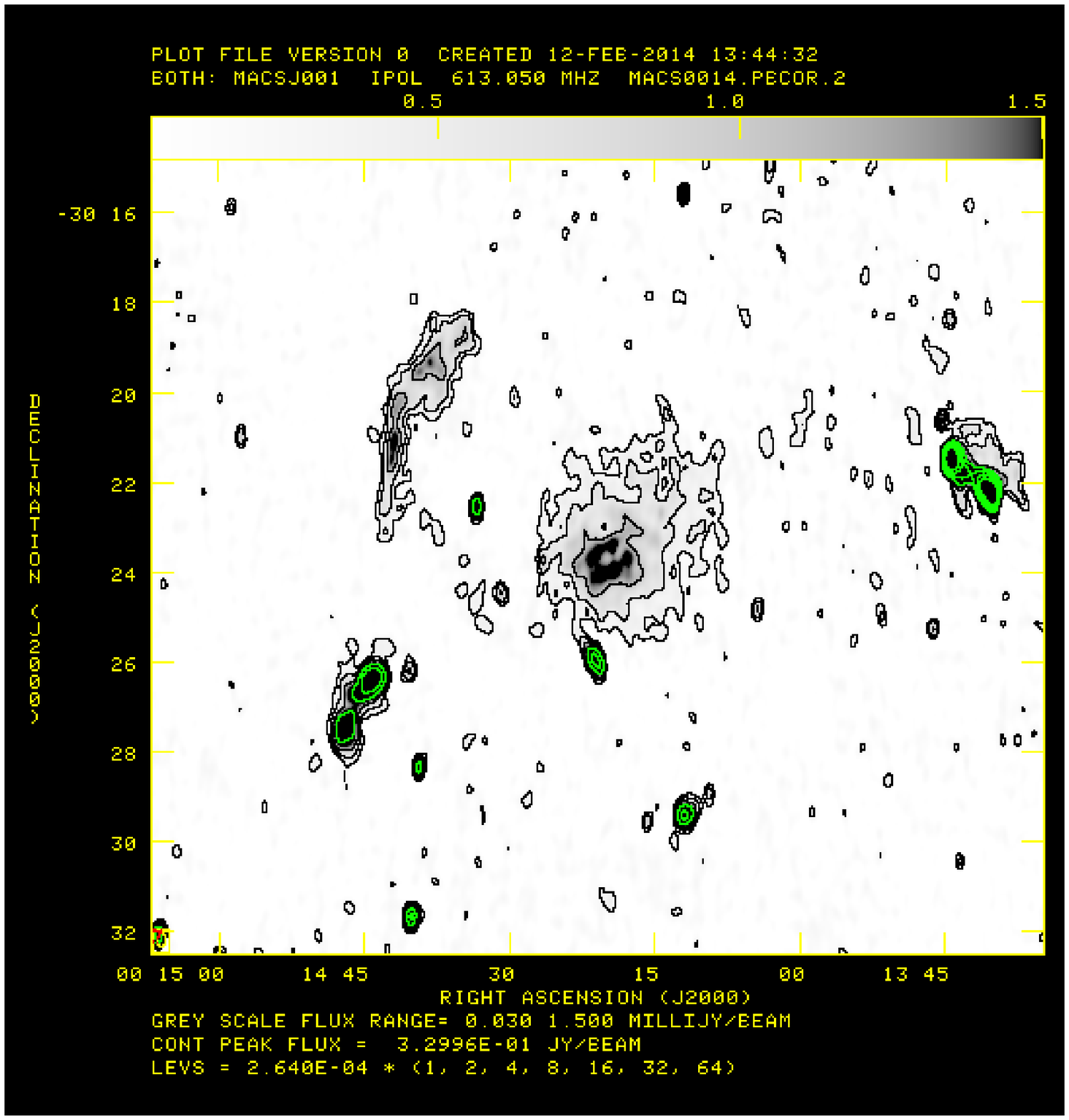}
  \includegraphics[width=4.5cm]{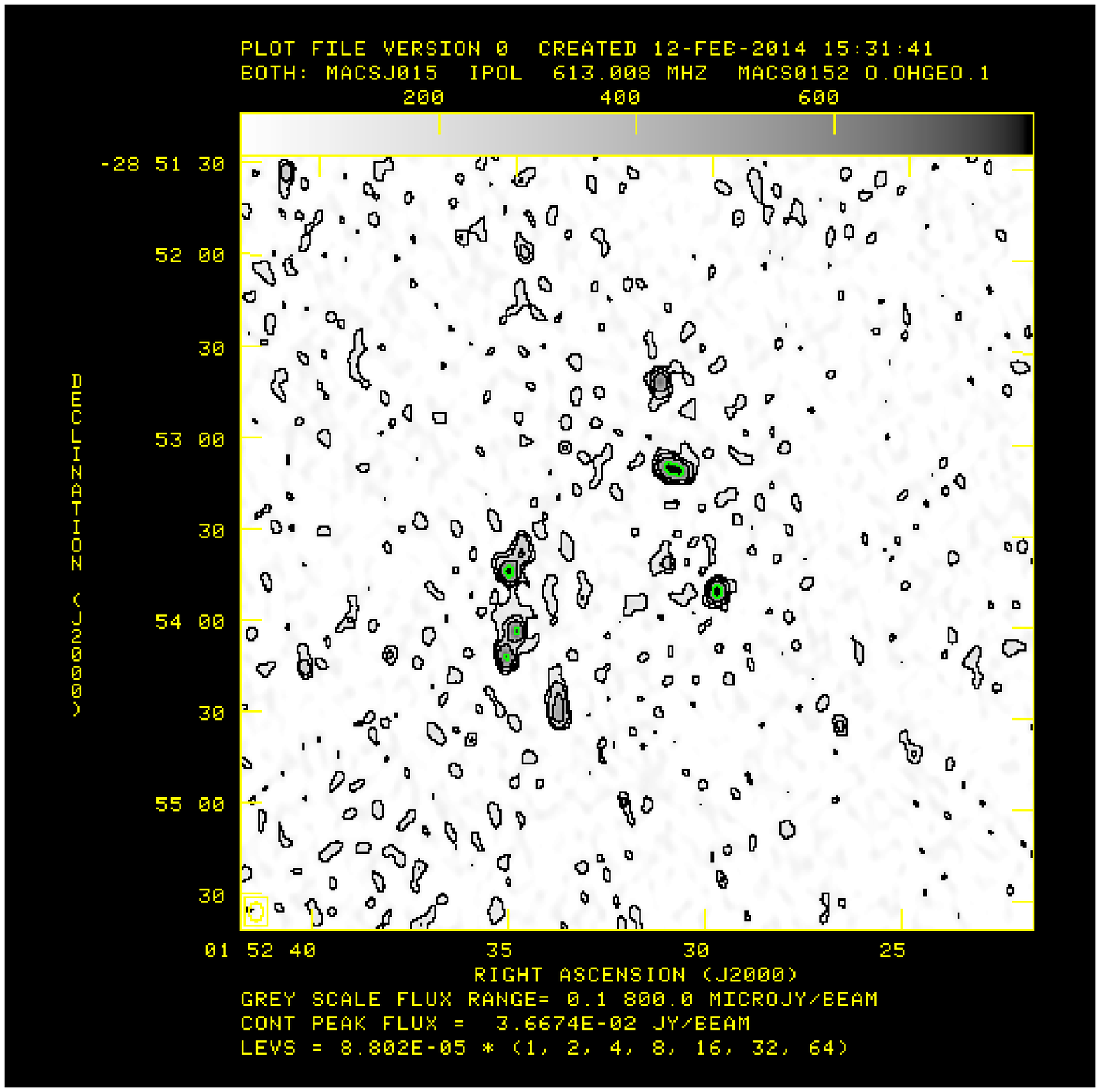}
 \includegraphics[width=4.5cm]{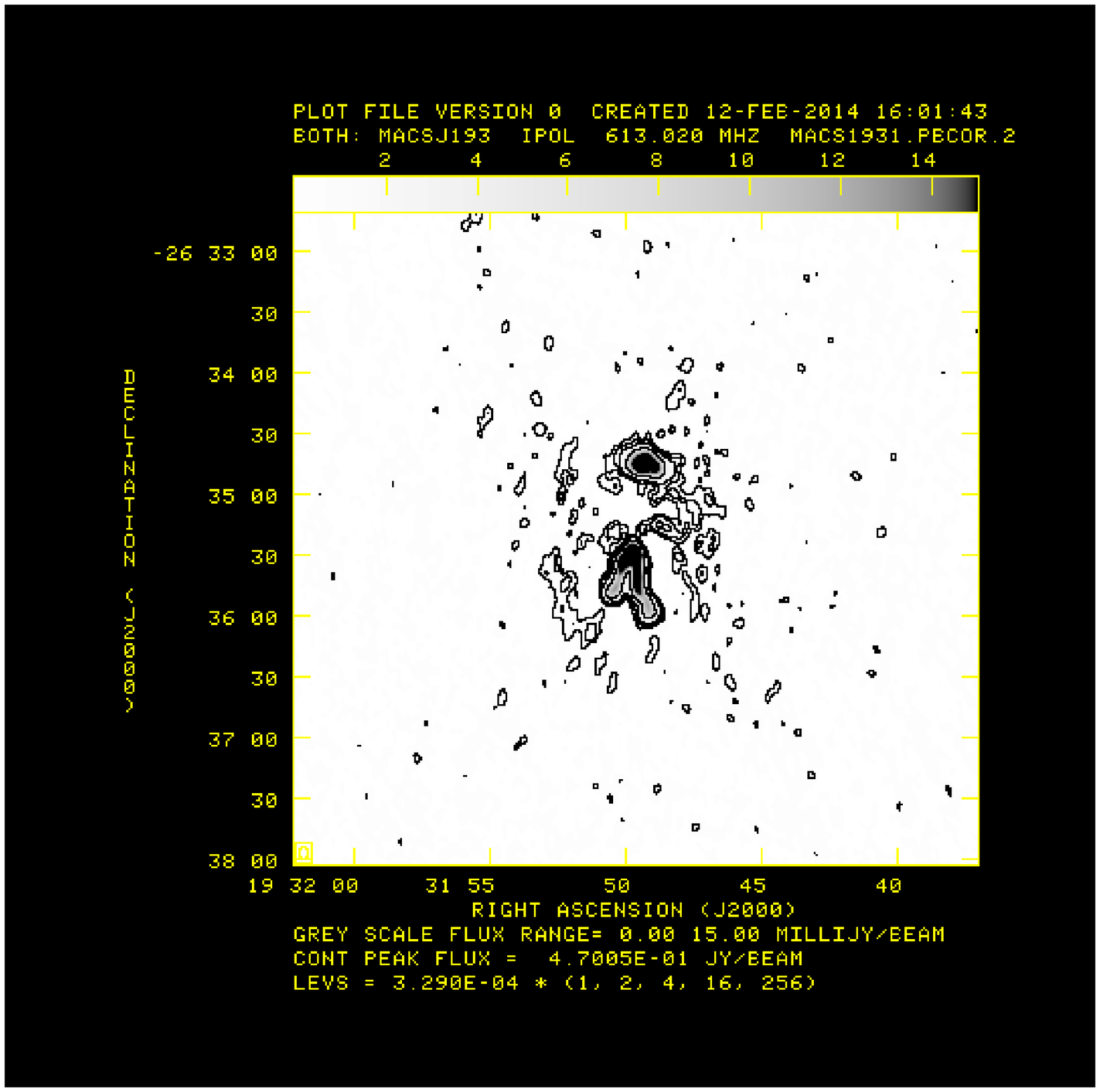}\\
 \includegraphics[width=4.3cm]{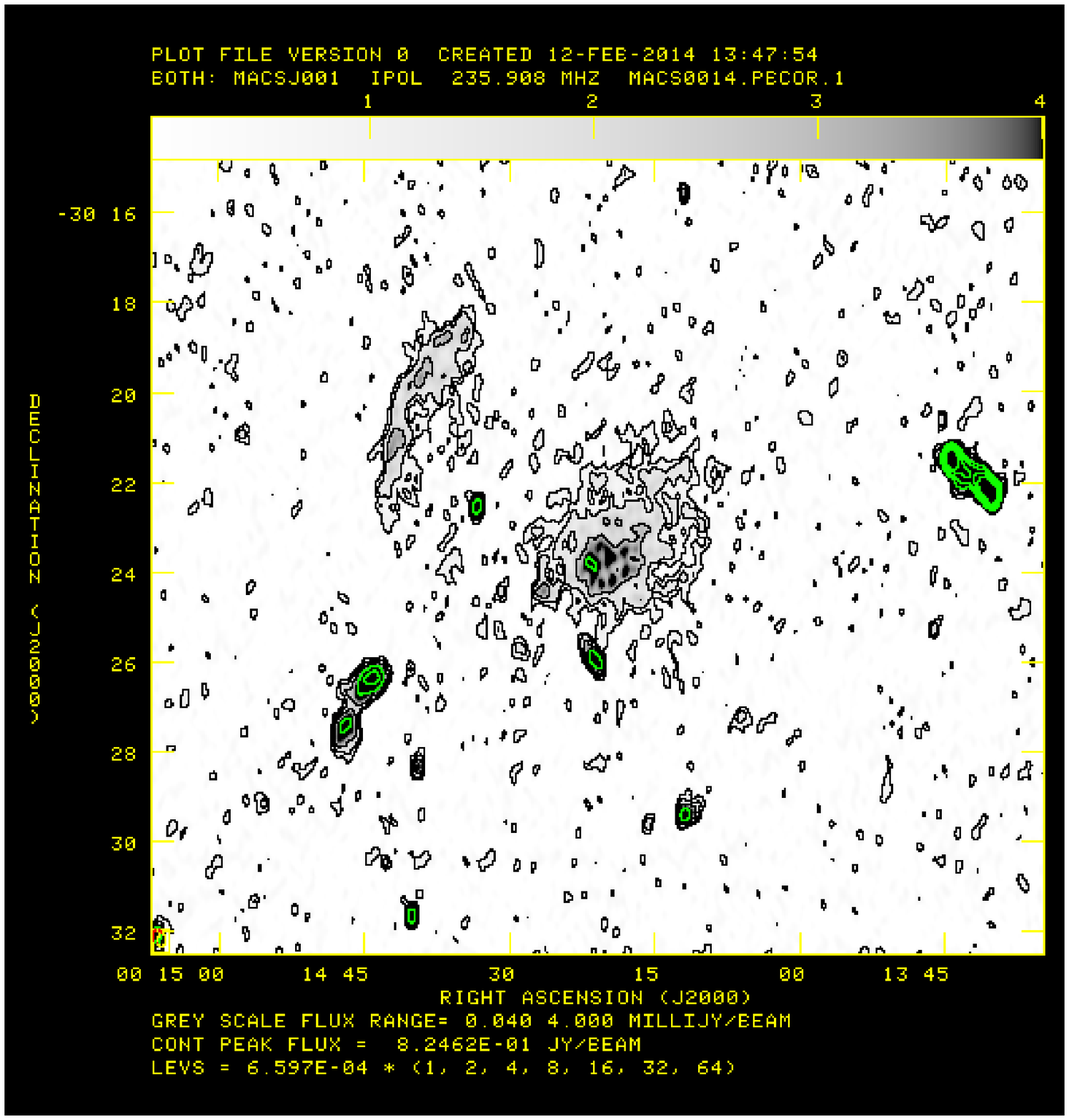}
  \includegraphics[width=4.5cm]{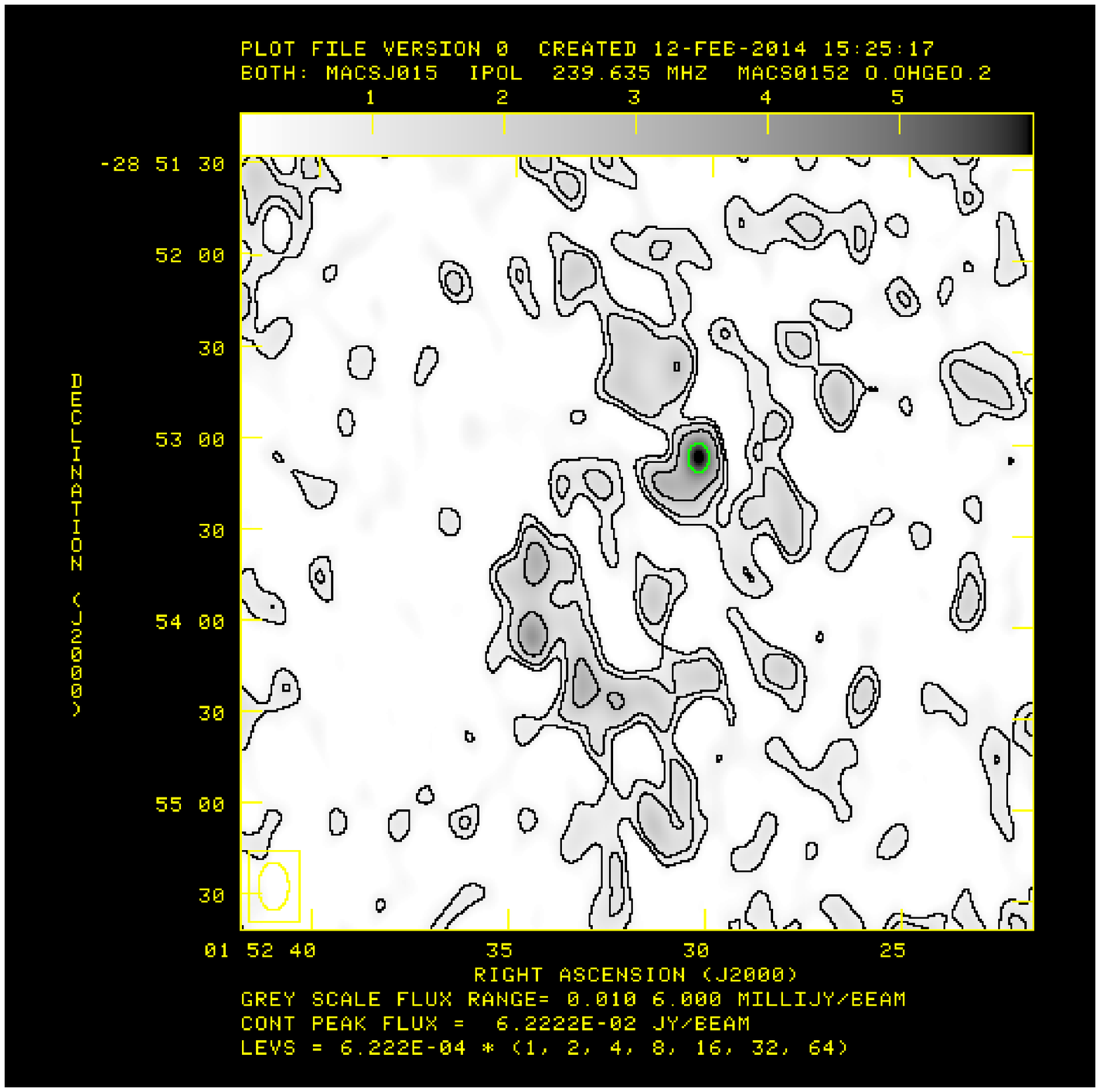}
 \includegraphics[width=4.5cm]{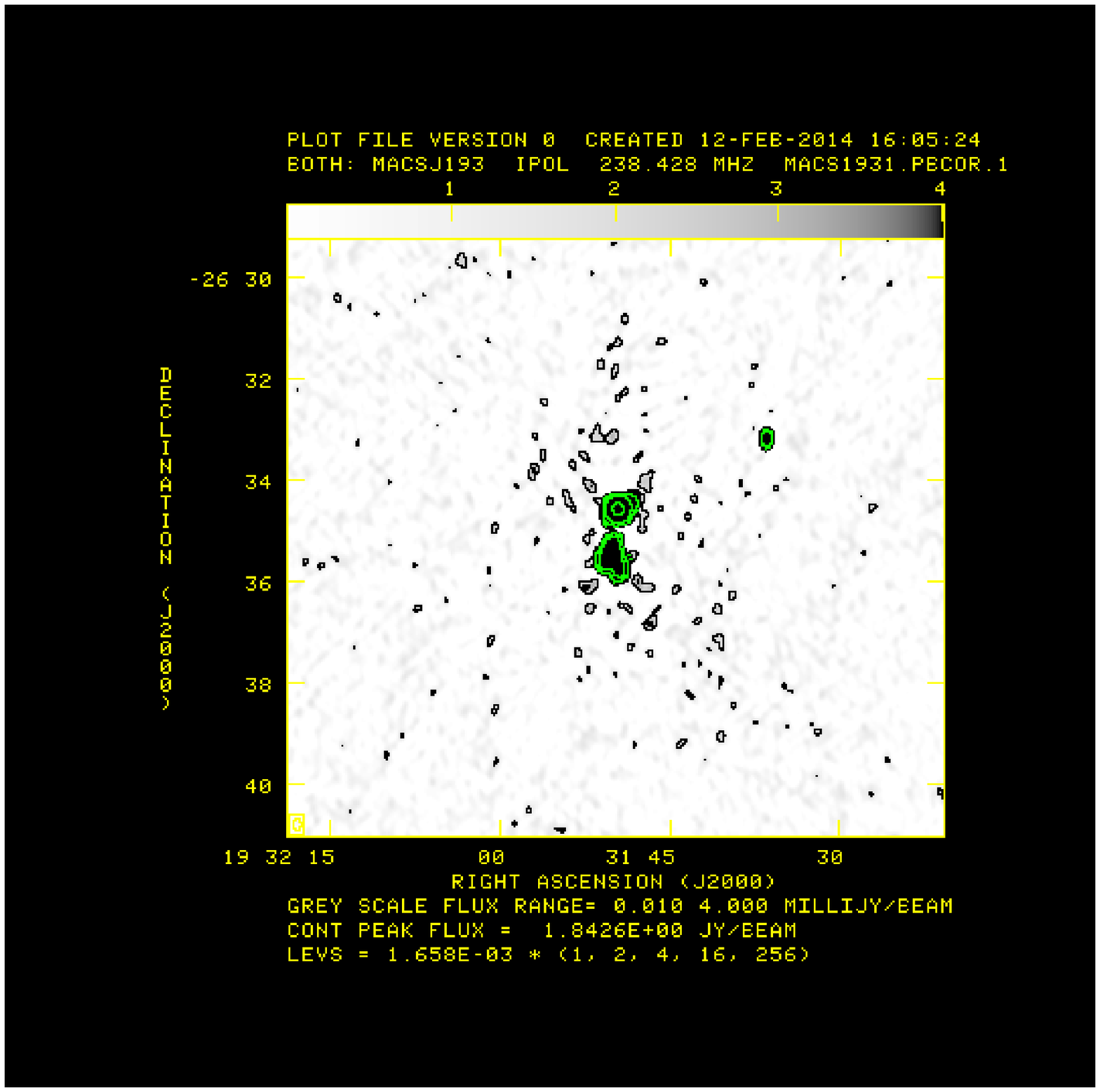}
\caption{ {\bf Top Panels:} 610 MHz GMRT radio continuum map as grey scale and black contours for the three clusters namely MACSJ0014.3-3022, MACSJ0152.5-2852 and  MACS 1931-2635 respectively. Specific sizes, contour level of flux are written on the images itself.  {\bf Bottom Panels:} GMRT 235 MHz Radio images with all other details as above.\label{radio}}
\end{figure}

\begin{figure}
\begin{center}
\begin{tabular}{p{6cm}cp{6cm}}
\raisebox{-\height}{\includegraphics[width=5.5cm]{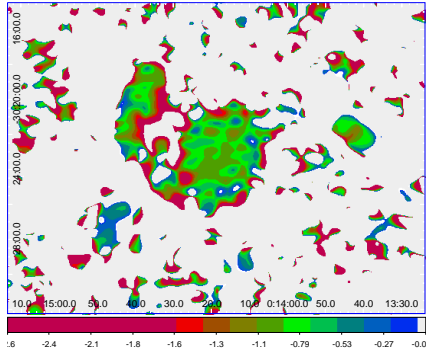}} & \quad &
\caption{This is the spectral index image of MACSJ0014.3-3022. GMRT 235 and 610 MHz images with synthesised beam width of $35^{''} \times 35^{''}$ are combined to construct spectral index maps of the east side relic and central radio halo.\label{f:spectral}}
\end{tabular}
\end{center}
\end{figure}

Interestingly, in MACSJ0014.3-3022, one of our four observed clusters, we have detected both a magnificent relic and an extremely large halo (Panel 1 and 4 in Fig~\ref{radio}). Both the relic and halo are almost of the same dimension of $\sim$ 1.5 Mpc. The relic is placed at more than a Mpc away from the cluster centre i.e. at the virial radius of the system. This object is also known as Abell 2744, with previously studied halo \& relic (e.g., \citep{Govoni2001,Orr2007}. Interestingly, it's also one of the HST frontier fields.
We made a spectral index map by combining GMRT 240 MHz and 610 MHz data (Fig~\ref{f:spectral}). Spectral index of the relic is steepening gradually from $\sim$ -0.7 at the outer surface to $\sim$ - 1.3 towards the inner side of the relic. This is because of higher shock compression rate and greater efficiency of particle acceleration through first order Fermi acceleration. On the contrary the halo has a flatter spectral index ($\sim$ -0.8) in the central part and it steepens towards the outer part of the halo ($\sim$ -1.1). The other interesting object is MACSJ0152.5-2852 (Panel 2 and 5 in Fig~\ref{radio}). This cluster is a very high redshift  z=0.413 cluster, with a possible relic of more than 0.5 Mpc long. Possibly, one of the earliest and young merging system detected in radio waves. The cluster MACS 1931-2635 (Panel 3 and 6 in Fig~\ref{radio}) doesn't show any clear diffuse emission. But, an interesting bent jet and a central bright radio galaxy is found. We didn't detect any significant radio emission from the cluster MACSJ0025.4-1222.

\section{Conclusions}

Our dual frequency (235/610 MHz) GMRT observations  have produced interesting results. We have discovered a unique object MACSJ0014.3-3022, with both a spectacular bow shock like flat spectrum radio relic and a huge central radio halo. This is also unique as, its central radio halo is unusually flat spectrum. Such a flat spectrum can only be possible if there is a continuous injection of shock accelerated high energy particles or  if ambient electrons are stochastically re-accelerated. So, this indicates the cluster is still going through a massive merging phase. 

\vspace{-0.4cm}
\section*{Acknowledgements}
{\scriptsize We would like to thank the staff of the GMRT that made these observations possible. GMRT is run by the National Centre for Radio Astrophysics of the Tata Institute of Fundamental Research. {\bf SP} Acknowledges DST-SERB Young Scientist funding under Fast Track Scheme (Govt. of India) for funding this project. {\bf HTI} acknowledges financial support by the National Radio Astronomy Observatory, a facility of the National Science Foundation operated under cooperative agreement by Associated Universities, Inc. {\bf AD} has been supported by NASA Postdoctoral Fellowship Program through NASA Lunar Science Institute.}
\vspace{-0.3cm}

\label{lastpage}
\end{document}